# Can ChatGPT pass a physics degree? Making a case for reformation of assessment of undergraduate degrees.

**K.A.Pimbblet**
**Centre of Excellence for Data Science, AI, and Modelling, University of Hull, Cottingham Road, Kingston-Upon-Hull. HU6 7RX**
**Email: k.pimbblet@hull.ac.uk**

**L.J.Morrell**
**School of Natural Sciences, University of Hull, Cottingham Road, Kingston-Upon-Hull. HU6 7RX**

**Abstract.**
The emergence of conversational natural language processing models presents a significant challenge for Higher Education. In this work, we use the entirety of a UK Physics undergraduate (BSc with Honours) degree including all examinations and coursework to test if ChatGPT (GPT-4) can pass a degree. We adopt a "maximal cheating" approach wherein we permit ourselves to modify questions for clarity, split question up into smaller sub-components, expand on answers given – especially for long form written responses, obtaining references, and use of advanced coaching, plug-ins and custom instructions to optimize outputs. In general, there are only certain parts of the degree in question where GPT-4 fails. Explicitly these include compulsory laboratory elements, and the final project which is assessed by a viva. If these were no issue, then GPT-4 would pass with a grade of an upper second class overall. In general, coding tasks are performed exceptionally well, along with simple single-step solution problems. Multiple step problems and longer prose are generally poorer along with interdisciplinary problems. We strongly suggest that there is now a necessity to urgently re-think and revise assessment practice in physics – and other disciplines – due to the existence of AI such as GPT-4. We recommend close scrutiny of assessment tasks: only invigilated in-person examinations, vivas, laboratory skills testing (or "performances" in other disciplines), and presentations are not vulnerable to GPT-4, and urge consideration of how AI can be embedded within the disciplinary context.

## 1. Introduction.

Education as we know it may well be dead. The release of ChatGPT heralds a new era in education that means the old ways of assessment must adapt or potentially become worthless or at best mistrusted if AI can successfully undertake such assessments (cf., Mahligawati et al. 2023; Polverini & Gregorcic 2024; Susnjak & McIntosh 2024; Yeadon & Hardy 2024).

Traditionally, assessment in undergraduate education is an important aspect of the learning process, as it helps to ensure that students are gaining the knowledge and skills they need to succeed in their chosen field and supports the consolidation and application of knowledge. The scientific literature on assessment in undergraduate education highlights several key principles and best practices that can help educators effectively evaluate the progress of their students.

The first key principle is that assessment should be closely aligned with the learning objectives of a course or program (Davies, 2019). This means that educators should carefully design their assessments to assess the specific knowledge and skills that students are expected to acquire (Biggs & Tang, 2007). This can help to ensure that students are adequately prepared for the challenges they will face in their future careers (Sluijsmans, Dochy, & Janssens, 2004).

The second important principle is that assessment should be ongoing and contain formative assessment as well as summative components (Black & Wiliam, 1998). This means that educators should not rely solely on traditional summative forms of assessment, such as exams and quizzes (Angelo & Cross, 1993), but should also use a variety of other formative (and summative) assessment tools, such as class discussions (McMillan, 2001), group projects (Angelo & Cross, 1993), and presentations (Black & Wiliam, 1998). This can help to provide a more comprehensive view of students' progress (McMillan, 2001) and allow educators to provide more targeted support and feedback (Angelo & Cross, 1993).

In terms of best practice, the literature strongly supports the use of authentic assessments (Palomba & Banta, 1999), which are tasks that closely resemble the challenges students will face in the real world (Angelo & Cross, 1993). These assessments can help to provide students with valuable hands-on experience (McMillan, 2001) and can help to improve their critical thinking and problem-solving skills (Palomba & Banta, 1999).

A significant fraction of the above – perhaps the dominant component depending on the exact domain in question – may be now at risk.

ChatGPT (https://chat.openai.com/chat) is a state-of-the-art natural language processing (NLP) model developed by OpenAI. It was released in November 2022 as a variant of the GPT-3 model, which at the time was one of the largest and most powerful language models in the world. ChatGPT is specifically designed to provide high-quality, human-like text generation for chatbot applications. It can generate responses to a wide range of input prompts, including open-ended questions, statements, and commands, making it a useful tool for building conversational AI systems. Open AI’s GPT technology has been shown to outperform other NLP models on various benchmarks and tasks (Radford et al., 2019), and has been used in a variety of applications, including dialogue systems, text summarization, and language translation (Brown et al., 2020).

One potential way that a student could use ChatGPT during their degree course is by using it to generate answers to (open) exam questions, assignments, or entire essays on a given subject (cf., Yeadon et al., 2023). By providing ChatGPT with a prompt containing the relevant course material, and the exam or assignment question, the student could generate a response that is highly likely to be correct and relevant to the topic at hand. Effectively the student could "outsource" their exam or assignment to ChatGPT, potentially allowing them to obtain a high grade without having to do the work themselves. In this way, ChatGPT can be used as a tool for plagiarism, allowing the student to pass off what the platform produced as their own (see, e.g., MacIsaac 2023; Roemer et al., 2024; López-Simó & Rezende 2024).

Using ChatGPT in this way raises significant ethical and academic integrity concerns, and educators and institutions will need to develop strategies for maintaining the integrity and value of their degree programs. We note here though that ChatGPT are not agents: they have no goal and as such are generating an output based on the input prompt – they are not necessarily “good” at physics or other disciplines, but they will have large training sets that are domain specific.

In this work, we apply ChatGPT to the summative assessment tasks from a physics undergraduate degree to see if it could pass, and if so with what grade. We undertake this primarily as an illustration of the ethical and academic integrity concerns that such technology raises. Previous work in the area such as Yeadon & Halliday (2023; see also Radenković & Milošević 2024; Kumar & Kats 2023) suggest a potentially weak performance in physics examinations – especially those that move away from fact-based recall and incorporated more open book responses with humans retaining some competitiveness (see also Susnjak & McIntosh 2024), but stronger performance in coding. It also goes further than previous work that use either previous versions of GPT or introductory courses (e.g., Kortemeyer 2023; Tong et al., 2023).

In Section 2, we describe the summative assessment of the degree in physics that we are seeking to pass with ChatGPT. In Section 3, we detail our approach, philosophy, and the dataset used. Section 4 presents our main results and findings, along with some detailed notes from individual modules that we looked at. We discuss our findings in Section 5 and present our conclusions in Section 6.

## 2. Dataset.

The dataset that we operate on is the University of Hull Bachelor of Science with Honours (BSc) in Physics as undertaken by students in the most recent iteration of the modules taught. This is chosen out of convenience for the authors in the main part, but also because due to the COVID-19, some examinations were taking place online at the time and therefore ChatGPT can retrospectively be tested in such an environment – assuming that it is not able to be detected; a point that we will return to in our discussions.

The BSc (Hons) in Physics at the University of Hull consists of six 20 credit modules per year, taken over 3 years that sum up to 360 credits for an award of a BSc (Hons). Different nations use different credit systems and we note here that 1 European Credit Transfer and Accumulation System (ECTS) credit is worth 2 UK credits for aid in conversions – see https://www.qaa.ac.uk/the-quality-code/higher-education-credit-framework-for-england . These modules, along with example topics, are given in Table 1. Each module has different assessment methods ranging from examinations through to coursework which have a variety of different weightings. All modules have a pass mark of 40% as an aggregate across all assessments. Students do not need to pass all assessment elements to pass a module, unless those elements are specifically designated as 'compulsory elements'. There is currently no provision for acceptable use of ChatGPT in the programme assessments, and thus its use would be considered academic misconduct. We use "cheat" as shorthand to describe this unauthorised use of ChatGPT constituting academic misconduct.

**Table 1. Summary of the BSc (Physics) module at the University of Hull. The right hand column gives GPT-4's score in the module along with any pertinent notes.**

| Higher Education Level | Module Name | Example Topics | Assessment Type | Assessment Weighting (%) | Assessment Mark | Final Module Mark (%) | Comments |
|---|---|---|---|---|---|---|---|
| 4 | Introduction to Experimental Skills and Mathematics | Laboratory skills Mathematics revision | **Mathematics assignment** | 50 | 91% | 55 | Fails compulsory laboratory element |
| | | | **Laboratory Skills** | 40 | 0% | | |
| | | | **Laboratory Calculations** | 10 | 90% | | |
| 4 | The Classical World | Classical mechanics Optics | Coursework | 40 | 89% | 57 | Pass |
| | | | Examination | 60 | 36% | | |
| 4 | Gravitation and Astronomy | Hubble law<br>Descriptive stellar evolution | Coursework | 40 | 20% | 45 | Pass |
| | | | Examination | 60 | 62% | | |
| 4 | Quantum Physics and the Properties of Matter | Introductory quantum physics<br>Thermodynamics<br>Crystal lattices | Mastering Physics | 40 | 80% | 57 | Pass |
| | | | Examination | 60 | 42% | | |
| 4 | Electricity and Magnetism | Programming in python<br>Classical electromagnetism | Examination | 50 | 38% | 63 | Pass |
| | | | Python task | 20 | 85% | | |
| | | | Python task | 30 | 89% | | |
| 4 | Experimental Physics and Mathematics I | Technical writing<br>Partial differential equations<br>Experimentation | **Mathematics assignment** | 50 | 90% | 60 | Fails compulsory laboratory element |
| | | | **Laboratory Skills** | 50 | 30% | | |
| 5 | Experimental Physics and Mathematics II | Continues from Experimental Physics & Mathematics I | **Mathematics assignment** | 50 | 95% | 55 | Fails compulsory laboratory element |
| | | | **Laboratory Skills** | 50 | 15% | | |
| 5 | Thermodynamics, Statistical Physics & Special Relativity | Ideal gases<br>Free energy<br>Length contraction<br>Time dilation | Special Relativity Assignment | 10 | 72% | 61 | Pass |
| | | | Thermodynamics Assignment | 10 | 65% | | |
| | | | Examination | 80 | 61% | | |
| 5 | Intermediate Quantum Mechanics with Advanced Computation | Angular momentum | Coursework | 20 | 100% | 93 | Pass |
| | | | Project | 30 | 100% | | |
| | | | Examination | 50 | 85% | | |
| 5 | Experimental Physics and Mathematics III | Continues from Experimental Physics & Mathematics II | **Mathematics assignment** | 50 | 92% | 55 | Fails compulsory laboratory element |
| | | | **Laboratory Skills** | 50 | 18% | | |
| 5 | Physics of Waves and Solid State | Crystalline lattices | Tutorials | 5 | 95% | 57 | Pass |
| | | | Assignment 1 | 20 | 96% | | |
| | | | Assignment 2 | 20 | 90% | | |
| | | | Examination | 55 | 60% | | |
| 5 | Stellar Structure and Evolution | Stellar interiors<br>Electron degeneracy pressure | Assignment | 25 | 100% | 90 | Pass |
| | | | Examination | 75 | 87% | | |
| 6 | Advanced Quantum Physics & Plasma Physics | Debye length Hamiltonian conjugates | Quiz | 25 | 70% | 61 | Pass |
| | | | Examination | 75 | 58% | | |
| 6 | Numerical Modelling and Simulation with Project Planning | Advanced programming Preparation for dissertation/research project | Programming assignment | 60 | 90% | 84 | Pass |
| | | | Literature review & project plan | 40 | 76% | | |
| 6 | Lasers and their Applications | Pumping levels Laser ablation | Class test | 25 | 67% | 78 | Pass |
| | | | Examination | 75 | 82% | | |
| 6 | Matter at Extremes | Superconductivity Particle physics | Class test | 25 | 91% | 85 | Pass |
| | | | Examination | 75 | 83% | | |
| 6 | Galactic and Extra-Galactic Astronomy | Galaxy structure AGN Cosmology | Coursework | 40 | 56% | 74 | Pass |
| | | | Examination | 60 | 86% | | |
| 6 | BSc Project | Open ended dissertation | Project report | 50 | Variable | Variable | Fails compulsory viva |
| | | | **Viva examination** | 30 | 0% | | |
| | | | Project skills | 20 | 0% | | |

**3. Approach.**

Our approach and philosophy to evaluate the effectiveness of ChatGPT is twofold.

Firstly, we adopt a "maximal intelligent cheating" approach. To us, this means using ChatGPT in an intelligent way to extract the maximum possible benefit from it in order to answer questions and assignments that might arise during a degree course where its use is prohibited. At minimum this means that we permitted ourselves to undertake the following actions where we deemed it appropriate and thought we could "gain" from it:

(a) Modifying questions for clarity in the hopes that a "better" answer might be produced or otherwise experimenting with the prompts given to ChatGPT;
(b) Choosing to split questions up into smaller (sub-)components and combining those components back together in a sensible manner where it is obvious that this could take place;
(c) Asking ChatGPT to expand upon sections of work (especially essay style questions) that it produced for us;
(d) Requesting references to the refereed literature or other publications where it was apparent it failed to include them during an earlier iteration;
(e) Using plugins (soon to be replaced by GPTs), coaching, and using GPT4's custom instructions where we feel it will help.

Secondly, we accepted the output to ChatGPT "as is" once we were content with the modifications and terms of use described above. In other words, we do not modify the outputs beyond what we are able to generate from ChatGPT. This second point is in direct conflict with the maxim of undertaking a maximal intelligent approach, since our intelligent cheat would take the output from ChatGPT as simply a starting point and try to go much further through, e.g., more standard internet search (noting that ChatGPT at the time of writing does not access the internet). Here we are simply interested in the ChatGPT output alone and therefore we go no further than its output. We note here that ChatGPT was prompted over a period of November 2023 through to February 2024. Image analysis and processing was not possible or implemented in ChatGPT during this period.

In undertaking the above, we acknowledge that we have a significant in-built advantage. By definition, as university teaching staff we possess insight that a typical student might otherwise lack. This, combined with the above principle of maximal intelligent cheating, likely means that the work presented herein represents a certain type of upper threshold that we can obtain from Chat GPT. We explicitly assume here that ChatGPT cannot be detected or evidenced sufficiently to support an allegation of misconduct – our cheating is assumed to adapt AI outputs that can bypass detection. The detection of ChatGPT use has been explored more widely in the literature (e.g., Rashidi et al., 2023; Elkhatat et al., 2023; Chaka 2023; Weber-Wolff et al., 2023; Zhang et al., 2024). We can summarize many of these studies by stating that the detection of ChatGPT is difficult to prove with certainty, and the false positive rate is significant – especially for non-native English speakers (cf., Liang et al. 2023).

Assessments were marked by the authors in accordance with the published marking criteria/mark schemes. Future studies could improve upon this work by having the original summative assessment authors undertake the grading, but this was not possible for the present study due to staff turnover and workload considerations.

## 4. Results.

We describe each module in turn herein and give some examples that we found interesting from the outputs obtained from ChatGPT for illustrative purposes. We explicitly refrain from publishing the entirety of our results and grading due to colleagues' wishes.

We note here that ChatGPT was prompted over a period of November 2023 through to February 2024. Image analysis and processing was not possible or implemented during this period.

**Introduction to Experimental Skills and Mathematics.** This module is one of a number that contain compulsory assessment elements (elements which must be passed to pass the module). Here, both the mathematics part (50% module) and the laboratory part (50% module) must both independently be passed for the module to be passed. The most immediate problem for ChatGPT is how it would handle being able to pass this laboratory component (or, indeed, any laboratory component of a lab-based degree) since it is not yet autonomous. For this particular module, 40% of the final grade is associated with undertaking measurements and instrumental competency, and experiments within the laboratory combined with uncertainty measures and propagation, plus reflecting on these. We assign 0% to ChatGPT for these since the lab is impossible for ChatGPT, but we do note that uncertainties and reflections are an area where it can perform well otherwise. A further 10% is associated with comprehension of laboratory safety standards and undertaking risk assessments. ChatGPT can formulate all but a full answer to this (9 out of the available 10 marks).

For the mathematics component of this module, ChatGPT performs better. There are a variety of take-home assignments for this component that test vectors, matrices, and calculus (11%), coupled with three papers conducted under exam conditions (13% each). Previous version of ChatGPT were unable to parse diagrams effectively, but this has more recently been rectified by it being able to read in figures in GPT4, an improvement on its capabilities at launch although still not perfect. Any question that has drawn questions (e.g., involving vectors pointing in certain directions) might not be read, and the technology is unable to provide an answer immediately. If we adopt the principle of maximum intelligent cheating, then we can describe to ChatGPT the angles that some of the vectors make with respect to the horizonal or vertical. Interestingly, ChatGPT makes good attempts at describing diagrams, and potentially drawing diagrams. When asked to sketch a diagram of 3 vectors, it readily produced complete asyptode code commands such as:

```
draw((0,0)--(0,-1), Arrow(6));
draw((0,0)--(1,0), Arrow(6));
draw((0,0)--(0,1), Arrow(6));
```

With this clue, we were able to turn diagrams into asyptode code and have ChatGPT interpret diagrams accordingly. For other questions and modifications to prompts, ChatGPT will produce ASCII art answers. The questions from the mathematics part of this module are answered almost entirely correctly otherwise, with only some diagrammatical questions proving stubbornly hard to parse to and from ChatGPT. This results in 91% for the mathematical element.

The overall grade assigned to ChatGPT for this module is therefore 55%. But it has failed the physics lab hurdle requirement and thus failed the module regardless. We will address how it might have passed in the discussion below.

**The Classical World.** In essence this module almost entirely about classical mechanics. Coursework accounts for 40% of the module, while a final examination accounts for the rest. Here, coursework is taken from an online tutoring and questioning system with a physics specialism. Lower-level questions (factual recall and simple calculations) are exceptionally simple for ChatGPT to resolve. As above, graphical questions are very hard to parse in the first instance, but with some descriptive effort on our part some can be answered correctly. In passing, we would note that this is probably one area where an internet search – especially of domain specific boards – may prove of higher utility. Regardless of this, ChatGPT obtains 89% of the available marks for the coursework element.

The examination proves to be a challenge in parts, but simple in others. For several straight forward questions, ChatGPT answers very incorrectly. One pertinent example that stood out during early (pre-GPT4) testing is as follows:

*Q: Consider the astronauts aboard the International Space Station at a distance of 420 km from the Earth's surface. Earth has a mass of $5.97\times10^{24}$ kg and a radius of $6.37\times10^{6}$m. The ISS has a mass of $4.20\times10^{5}$ kg. What is the magnitude of the acceleration due to gravity for the astronauts?*

The answer given here is entirely correct up until the last possible moment, wherein ChatGPT gives an incorrect answer of $9.81\text{ms}^{-2}$ (rather than the correct $8.64\text{ms}^{-2}$)

To understand why ChatGPT had given this incorrect answer, we instead asked it to "show" that the correct answer should be $8.64\text{ms}^{-2}$. It readily came to the right answer when the prompt "show" was used instead of the wording of the question above. On asking why it had initially given an incorrect answer, ChatGPT responded that "The first answer was given as $9.81\text{ms}^{-2}$ because it is the standard value for the acceleration due to gravity at the Earth's surface. This value is often used as a default value or rough estimate for the acceleration due to gravity in many situations, as it is relatively easy to remember and use." So even though we feel this was a simple question that it should have got correct, it is clear that ChatGPT can over-ride itself to give "default" answers to technical questions. Indeed, this is seen in some questions where it defaults to more accepted answers for simple questions. This effect is much less pronounced in more recent versions of the software (cf., ChatGPT-4o). More descriptive answers were generally done well, including describing why astronauts are considered “weightless”. Finally, ChatGPT fails regularly with multiple step solutions that are required for many questions – even version 4, a pattern that is repeated across multiple modules. From experimentation, we can sometimes get ChatGPT to output the answer to an early step in the question and if we are able to coax this out, then the subsequent part(s) become more likely to be correct. However, in the main part, questions that require deep analysis and multiple steps to obtain a correct answer are more likely to be failed, albeit with some grades given for the early steps.

For questions that were of a more descriptive nature (e.g., “How would the cork ball [move through the water] if it was a different size?”), ChatGPT performed more strongly and gained more marks from the examination. Overall, ChatGPT obtained 36% for the exam – a failing mark. Combining with the coursework, the overall grade for the module is 57% as there are no compulsory elements in this module.

**Gravitation and Astronomy**. This module has both a coursework component (40%), and a final examination (60%). The coursework component asks students to plan and execute an astronomical observing run. This uses knowledge about right ascension and declination, alongside undertaking the observations. Naturally, ChapGPT cannot undertake the observations, and generally fails to produce adequate explanations about whether certain (Right Ascension, Declination) coordinates will be visible on given nights as this is not necessarily an easy task to undertake. Indeed, it recommends that we use specialist astronomy software to undertake the task instead of asking itself. It scores 20% on this coursework.

The examination fares better, although with the caveat about pre-GPT4 versions being unable to parse diagrams from questions. There are plentiful questions that “describe” the Universe in general terms and ChatGPT does very well indeed on these. Even questions where the current values of Universal constants are deliberately chosen to be fake, ChatGPT is able to explain what might happen next and make adequate predictions. We found a case as well where ChatGPT decided to use Newton’s version of Kepler’s third law, where the latter would have been more appropriate and simpler to implement. It still got the correct answer. More curiously, ChatGPT performs notably better at multi-part solutions to questions than it did with classic mechanics in *The Classical World*. We guess and infer that there might be either more astronomy material in its training set, or there are fewer permutations of introductory astronomy questions than classical mechanics ones. Overall the examination scores 62%. Combining this with the poorer course work results in a final grade of 45% which although a pass mark, reflects the poor performance in the observing coursework.

**Quantum Physics and the Properties of Matter**. The coursework is from an online physics tutoring system which ChatGPT scores well on (scoring 80%). The examination component is split between the two topics in this module. ChatGPT performs very strongly on the quantum physics component (25/30), but very poorly on the properties of matter component (5/30) which yields 42% overall for the examination. Part of the poor performance here is once again the inability to parse graphical questions. However, even though ChatGPT refused to even attempt to sketch a graph for several questions, it did describe the graph well. Posed with the question “For a particle in an infinite potential well of width $L$, plot the wave function and the probability density for $n = 2$.”, it responded by describing the curve very carefully including where x=0 would be located. We were puzzled by the inclusion of the Lande factor, g, when asked to undertake a magnetic field derivation as it is not a regular feature at this level in UK HE, although may be more commonly taught and used elsewhere. A final grade of 57% is obtained for this module.

**Electricity and Magnetism with Computation**. Split evenly between programming in Python and classical electromagnetism, this module has three distinct assignments including an examination on the latter (50% weighting) and two python programming assignments (20% and 30%) that represent a variety of relatively straightforward python tasks. As an example of the computation tasks, one of them is *"Create a Python function that evaluates the Shockley diode equation for an input numpy array containing a sequence of potential difference values. Other inputs should allow for the other non-constant variables to be passed to the function and the resulting current should be returned.*" ChatGPT performs superbly well at addressing these questions and sometimes goes beyond standard answers, in particular by providing extended explanations of what it creates using comments. Although imperfect in places, the assignments are scored at 85% and 89% respectively. Meanwhile, the electromagnetism exam exhibits now-familiar issues. Interestingly, it can cope with derivations very well, but struggles once more with multi-part questions and tests of reasoning and logic. The score for the examination is 38%. This results in an overall grade of 63%.

**Experimental Physics and Mathematics (HE4, HE5).** The three modules at HE4 and HE5 entitled *Experimental Physics and Mathematics* follow a pathway through increasingly complex topics in the subjects. These relate to the types of experiments conducted and their length, experimental design, scientific writing skills (HE4), differential equations, Fourier analysis, through to vector calculus and La Placians. Both aspects (experimental skills and mathematics) of the module must be passed for an overall pass, although there are a range of different assessment across the three modules, including portfolios of work on the experimental side and multiple problem sheets and examinations on the mathematical side. As in *Introduction to Experimental Skills and Mathematics*, the inability of GPT to physically enter the laboratory is a hurdle that results in an overall fail mark. However, there a lot of capabilities that GPT-4 is good at, ranging from handing uncertainties, through to considerations of health and safety and creating risk assessments documents in general terms. For the mathematics, it performs strongly in almost every aspect. The weaker areas are more in-depth questions that require multiple stages, and diagrammatic answers which although it sometimes attempts, is generally poorer at. The comments noted above thus remain valid and we simply note here that the grades overall are typically 50-60% for the module in question (we will average this to 55% in our final analysis), but GPT fails the compulsory laboratory element once more.

**Thermodynamics, Statistical Physics, Special Relativity.** A suite of thermodynamical concepts is covered in this module ranging from the laws of thermodynamics through to statistical mechanics and advanced concepts in entropy and engines. Assessment is from a pair of assignments (one for each of thermodynamics and special relativity) weighted at 10% each, and a final exam (80%). GPT4 performed exceptionally well on the special relativity assignment, including a lot of detail – most especially for counter-intuitive results such as where a rod is length contracted sufficiently to fit in a container that it otherwise wouldn't. Thermodynamics is performed at a strong level as well, but not quite as strong as special relativity in general. We were impressed with GPT4 using (unprompted) Stirling's Approximation to help simplify factorials in one of the answers. One "proof" style question that it struggled with was the derivation of Helmholtz free energy from a knowledge of the partition function. This was unexpected, but likely due to going down a rabbit hole of incorrect working, and possibly due to inputs not being quite in a format it could deal with (most especially large brackets). Overall, GPT4 performed strongly enough in the module to achieve 61%.

**Intermediate Quantum Mechanics with Advanced Computation**. On the computational side, ChatGPT once again performs superlatively well in both the in-semester assessment (20%), and a longer term in-semester project (30%), both based around quantum mechanics and modelling particles in various scenarios such as the creation of a Hamiltonian matrix. We feel justified in giving ChatGPT 100% for this component. The rest of the grades (50%) arise from a quantum mechanics examination. GPT4 shows remarkable insight into the examination questions, including typical potential set ups such as an electron having energy E incident on an infinite width potential step of height V (where the potential is zero for x<0 and V for x>0) and the energy is always less than V. We were impressed that even with this very simple description, ChatGPT volunteered the boundary conditions and solution to Schrodinger's equation in both regions, along with consideration of boundary conditions, transmission, and a physical interpretation without additional prompting – which essentially covered several of the subsequent sub-questions. Remarkably, the performance of GPT4 on the examination is exceptionally strong with an 85% grade leading to an overall module grade of 93%.

**Physics of Waves and Solid State.** There are a series of mini-tutorials associated with this module that comprise 5% of the overall grade. These feature shorter questions to ensure students have grasped basic concepts. ChatGPT performs superlatively well on these questions and achieves almost 95% with a one-shot approach and little need to break down questions into smaller components. Some minor mistakes were made with parsing a diagram featuring conduction and valence bands with a note of where the Fermi level resided. Two further assignments (one on waves, and one on solid state) are worth 20% each, and a final is worth 55%. As previously, GPT4 struggles with some of the more diagrammatical questions, but is certainly more than capable of describing an output. We explicitly note here the description given of how to sketch a Face-Centred Cubic (FCC) lattice given in Figure 1.

1. **Start with a Cube**: Draw a cube. This will be the basic framework for placing atoms at the corners and face-centers.
2. **Corners**: Mark points at each corner of the cube. In an FCC lattice, atoms are located at every corner of the cube.
3. **Face-Centers**: On each face of the cube, mark a point at the center. These represent atoms that are positioned at the center of each face of the cube.
4. **Visualization**: Now, imagine or draw lines connecting these points to visualize the close-packed planes. The FCC structure is characterized by atoms being packed as closely together as possible, with each atom in the structure touching twelve neighbors: six through the face-centers and eight through the corners (since corner atoms are shared among eight adjacent cubes).

Figure 1. Output from ChatGPT when prompted to give a description on how to sketch a FCC lattice.

Similarly, a sketch of the Hall Effect was attempted within GPT4 as shown in Figure 2.

**Sketch Description:**

Imagine a rectangular conductor placed in the xy-plane, with a length (L) along the x-axis, width (W) along the y-axis, and thickness (d) along the z-axis. An electric current (I) flows along the x-axis, and a magnetic field (B) is applied along the z-axis, perpendicular to the current. Due to the magnetic field, charge carriers (electrons for a typical metal) experience a force that pushes them towards one edge of the conductor, leading to a voltage (V_H) across the y-axis (width of the conductor). This voltage is known as the Hall voltage.

javascript Copy code

```javascript
B
↑
|    I →
|  /_______
| /       /|
|/_______/ |
|       |  /
|       | / V_H
|_______|/
```

Figure 2. ChatGPT’s response to sketching the Hall Effect.

In the assignments, GPT4 scores 96% and 90% respectively. For the final examination, GPT4 performs better with the solid state component than the waves component, and scores 60% overall. This yields a 2:2 level pass at 57% overall.

**Stellar Structure and Evolution**. This is a classical course in stellar structure that features common content in terms of differentials, electron degeneracy, and general evolution of stellar structure. The summative assessment consists of an assignment worth 25% and a final examination worth 75% of the module, supported by formative tutorial questions every week that help scaffold understanding. We focus only on the summative assessment here. GPT4 scores a very note-worthy 100% on the assignment that consists of a balance of written work and numerical work. We assume that the style of questions used here commonly used, or are otherwise “well known” in the education community. The exam is similar, with a score of 87%. The wrong answers are strong variants of known derivations and questions with unique circumstances applied. Overall, this is a module passed with a mark of 90%.

**Advanced Quantum Physics and Plasma Physics**. From Debye lengths through to Hamiltonian conjugates, this module is the senior level course for quantum mechanics that brings into play relevant plasma physics. The assessment consists of formative tutorials, a summative quiz (25%), and a summative examination (75%). In this regard, the module stands out as lacking in coursework elements and is wholly test (performative) based. The quiz is answered very well in the main part with only the diagrammatical questions posing a challenge for GPT4 to correctly parse. A score of 70% is obtained here. The examination is more mixed. In attempting to answer a question about the current in a plasma using Ohm's Law, GPT4 reverted to the more usual V=IR formulation rather than appreciating to use J=sigma E. This likely attests to the former being vastly more common in the training set and might provide a pathway for examiners to consider using more advanced versions of common equations to deceive GPT4 under some very specific conditions. In using the Appleton-Hartree formula, there was also confusion between N and n (refractive index, and electron density) which we found interesting. This might also reflect a confusion in the training set wherein it is plausible that inputs might incorrectly capitalise the letter (or not). This again represents a possible method for examiners wanting to proof against GPT4, but we do not claim to have pushed the boundaries of this hypothesis in the current work. An score of 58% is obtained for the examination which results in a final grade of 61%.

**Numerical Modelling and Simulation with Project Planning**. In this module, students refine their Python programming skills and apply it to a specific assignment context (e.g., the Sun's wobble or a simulation of a binary system) which is worth 60% of the module. In general, ChatGPT excels at this task. We suggest that the reason for this is that this is already a "solved problem" with a good number of different implementations present on the internet. Although we did not check code similarity to other sources, the code produced works, even if it is not the most efficient. GPT4 scores 90% on the assignment. The Project Planning element of the module is a slightly different tale. Here, students undertake a literature review and plan the research project that they will undertake later. The difficulty with this assignment is, of course, the open-ended nature of the project that they will undertake. Students are given a selection of project topics and short descriptions to select between as a starting point and subsequently engage with relevant members of staff to discuss. We attempted to replicate three different submissions by students spanning the range of length scales of the Universe. In general, GPT4 was well able to produce literature reviews for the project and make sane and sensible suggestions for the planning component – we explicitly note that ChatGPT is well-able to provide non-hallucinated sources and the prospects for identifying AI-generated text in this manner when prompted appropriately are minimal. We award GPT-4 an average of 76% for the project planning resulting grade for this module of 84%.

**Lasers and their Applications** includes topics such as Einstein coefficients A and B, pumping levels, through to laser applications in industry such as ablation. Assessment is entirely by examination with both a class test mid-way through (25%), and a final examination (75%). The class test is conducted online and features parsing of graphical information that GPT-4 performs poorly at. However, the longer more descriptive questions are answered almost perfectly (e.g., "Write a short account of how axial mode beating can be used to produce short laser pulses"), and the assessment scores 67%. The final examination is extremely similar in execution with the graphical answers being tough to parse, but the descriptive and numeric answers very simple to solve. In Figure 3, we show a diagram outputted as part of a question about a 4 level laser system that GPT-4 used to answer a question asking it to explain both 3 and 4 level lasers and how they worked in order to compare and contrast efficiencies. The score for the examination is 82% which yields 78% overall for the module – a first class outcome.

**Energy Diagram:**

```scss
E4 (Pumped State)
  |
  | (fast non-radiative decay)
  V
E3 (Metastable State) ------------> Laser Emission
  |                                  |
  | (stimulated emission)            |
  V                                  V
E2 (First Excited State)
  |
  | (fast decay to ground state)
  V
E1 (Ground State)
```

Figure 3. Diagrammatical output sketch from ChatGPT to explain 3 and 4 level lasers.

**Matter at Extremes.** Discussing a range of physics from particle colliders through to superconductivity, this module touches on a variety of rarer phenomena in the discipline and examines students through a small class test (25%) and a large final examination (75%). Both components score strongly in straight forward computations and discussions of physics. We were impressed by the quality of answers here but are are unclear why this might be the case – perhaps a limited set of questions (or variations thereof) exist, or perhaps the nature of the questions is fortuitously coincidental with GPT-4's training set. However, the same diagrammatical issues plague responses here as noted previously. In addition, we note that some of the equations and terms were represented as images (as opposed to text in the pdfs that we copied and pasted the questions from). These caused issues with transposition into GPT-4 and, in extremis, may constitute a "frustration" level that students could determine it were easier to tackle the questions themselves than ask GPT-4 to do so. The final grade here is a first class 85% overall.

**Galactic and Extra-Galactic Astronomy**. This module goes from the Milky Way and its structure all the way out to cosmology and the Friedmann equations that govern the evolution of the Universe. By necessity, there is a lot of knowledge to be learned within this module, but the examination (60% of the final grade) focuses not simply on repetition of this knowledge, but novel applications. The coursework (40%) is a roleplay scenario wherein students are asked to respond to an imaginary email asking for their help in designing a computational simulation. The examination is performed well, with even questions that require diagrammatical outputs being tackled with ease, albeit descriptively. As noted in some examples above, GPT-4 is not able to produce sketches, but it does a terrific job at instructing the user how to produce such sketches (e.g., of the structure of the Milky Way). In this manner, it may be impossible to detect a student using GPT-4 since although the instructions will be correct, each student sketch will by definition result in a variable output. The exam scored 86% overall. The coursework is less well performed, although correct citations are used for it and some of the explanations used are strong resulting in 56% in this component. The overall final grade for the module is 74%.

**B.Sc. Project.** Running a dissertation project of this nature through GPT-4 represents a tremendous problem for this analysis. The first major problem is one of diversity. In principle, we could simply attempt to replicate a project – arguably a computational one for which GPT-4 arguably may perform well. However, project module is also assessed by an individual viva (oral examination), and a project skills mark awarded in part by the project supervisor for performance during the project process. Regardless of how well GPT-4 does on the final written project, it is very clear that in all cases it would fail the viva for the project and as this is a compulsory element which must be passed, we make the argument here that the project will simply be a fail overall due to the nature of the viva examination for the project component.

It should be noted that there exist other technologies that may facilitate the use of AI during oral examinations. The ability to have a viva conducted remotely and have AI work in real time to address questions live during the viva could plausibly result in a pass. Such technology (co-piloting) is already being deployed for remote interviews, for example. However, for more complex cases such as a viva based on a technical piece of work, this is more advanced and beyond the scope of the current work given where GPT-4 is currently at.

## 5. Does GPT-4 Pass the Degree?

To pass the degree, GPT-4 requires 40% overall as a weighted average. The weightings are such that the first year does not count toward the final degree classification and as such is a "qualifying year" wherein a student simply needs to pass to stay on the degree. The percentage scores for the Level 4 modules noted above are: 55, 57, 45, 57, 63, 55. This results in an average of 55%. However, GPT-4 has already failed the degree since as we note above, it cannot pass the laboratory hurdle, and so would not progress to the next stage of the degree programme. Our "maximally intelligent cheat" might therefore focus on the hurdle requirements for the laboratory assessments, participating in laboratory sessions while relying on GPT-4 to complete other assessed work.

The second year is weighted at 30% of the degree and the third year at 70%. In second year (ignoring the fact it fails once again in the laboratory) an average percentage of 69% is obtained with superlative performances in stellar astronomy and programming. Once again though, the laboratory situation would prevent the student progressing.

In third year, the average from 5 modules that have been passed is 76%. The final project was a failure due to the viva. If this were to happen in real life, then a student would be offered a re-sit in the first instance, where continued reliance on GPT-4 would result in a further fail mark. The University of Hull Regulations could allow the module to be "condoned" and allow the student to graduate, given that it is the only module failed, and that original work has been produced in part of the project planning and part of the laboratory ensures that programme outcomes are satisfied. Thus, we will award zero for the project and arrive at a final weighted average of 64% for the third year.

Overall then, GPT-4 obtains a final rounded grade of 65%. In UK terms, this equates to a 2.1 degree (upper second class honours), which offers better employment prospects with higher postgraduate earnings (Britton et al 2022). It cannot pass the degree due to compulsory laboratory elements. Hence the answer to our initial question is "no". ChatGPT (GPT-4) alone cannot pass the physics degree. It can, however, achieve good quality results on many types of assessments, and so could form part of a successful strategy for our “maximally intelligent cheat”.

Finally, we note that the advent of image processing and interpretation capability by ChatGPT could enhance the grades that it can score, although the extent to which this will help is not clear (cf., Polverini & Gregorcic 2024).

**6. Discussion.**

Although GPT-4 alone does not pass, we are able to draw some interesting generalities here which may have influence for practitioners about the abilities of GPT-4 at the time of writing. The tasks that GPT-4 performs superlatively well at include computations (programming up to a moderate level, and sometimes even more complex), and simple calculations and scenarios. As tasks become more complicated, or multi-step solutions are required, performance drops off. Written prose covers a range of performance and we find that multiple threads and divide-and-conquer approach to writing works very well. Parsing of graphical questions is a mild problem that we expect improvement on in the time ahead – hence we provide the obvious caveat here that everything we have tested is correct at the time of writing, but we expect everything to improve over time given current trends. In Table 2, we present a summary of performances based on different tasks, rather than sub-discipline as presented in the previous section. These outcomes are in broad agreement with question types probed by others (cf., Newton & Xiromeriti 2023) and it must be noted represents a strong advance over GPT-3 models (e.g., Gregorcic & Pendrill 2023; Ramkorun 2023; Tong et al., 2023; West 2023; Yeadon et al. 2023; Yeadon & Hardy 2024; see also Zollman et al., 2024).

**Table 2. GPT-4 performances. Percentage grades are given for the equivalent UK classifications. We note that a “good degree” is regularly taken to mean an upper second class degree or better within the UK and such grades are generally required for access to “graduate jobs”.**

| Percentage grade | UK Classification | Task types |
|---|---|---|
| 85-100% | High First Class and near-perfect scores | Multiple choice questions.<br>Generic coding tasks. |
| 70-85% | First Class | Single-step problems.<br>Some specific coding tasks. |
| 60-69% | Upper Second Class | Some short response prose.<br>Some two-step problems. |
| 50-59% | Lower Second Class | Some short response prose.<br>Some long response prose. |
| 40-49% | Third Class | Some long response prose.<br>Some multi-step problems.<br>Problems requiring insights.<br>Inter-disciplinary problems. |
| Less than 40% | Fail | Laboratory/practical skills.<br>Problems requiring deep insights.<br>Complex diagrammatical questions<br>Vivas.<br>Invigilated examinations. |

Of course, we have created a “false” situation here in some ways. We have used GPT-4 against examinations that would ordinarily be run in an invigilated format in our context. This is not always the case as was discovered during the COVID-19 era when all teaching pivoted to an online and generally asynchronous format. There is a case to be made here that open book examinations taken at home (for example) are highly vulnerable to unauthorised use of GPT-4. This also applies to coursework in general: almost any imaginable topic short of things requiring deep insight are vulnerable – and perhaps some aspects of original research as well (cf., Bubeck et al. 2013). This makes the case unambiguously for reconsidering what assessments are being used during a degree.

Two broad paths are open: the first is to ensure that assessments are robust to the vulnerabilities posed by GPT-4 such that our “maximally intelligent cheat” would be unable to achieve the level of success outlined above. The second is to embed these tools into the disciplinary context, supporting students to use them ethically, critically, authentically and with integrity (Beckingham et al 2024). Broadly, these align to the ‘avoid’ and ‘embrace and adapt’ strategies identified by JISC (2023). What is clear is that “business as usual” cannot continue.

**Path 1: Robust assessments:** To categorically ensure validity within current frameworks, then the unfortunate recommendation is that in-person assessment tasks (invigilated examinations, vivas, field work, lab skills, or “performances” in other disciplines, and perhaps aspects of oral presentations) are now a necessity. We have deliberately avoided “reflective” work in this list since with a reasonable prompt and access to data, even a reflective piece of writing is straight forward to produce. This is not to say that the noted forms of assessment should be used everywhere – although this may have been the expectation mere decades ago. Rather, strategic use of such assessment at the terminal ends of every year would be good to implement (e.g., end of year invigilated examinations), and an in-person viva for projects appear to be non-negotiable in order to defend outcomes. This is in agreement with Kortemeyer & Bauer (2024) who underscore the requirement for “high-stakes examinations in supervised settings to ensure academic integrity”.

This approach thus assumes that the unauthorised use of GPT-4 and similar tools constitutes academic misconduct, which raises a real issue with detection. While detection tools have been shown to work effectively against GPT-3.5 (Walters 2023), disguising GPT-4 output is relatively simple in the current era and the random nature of the outputs from ChatGPT is merely the first step in this. We posit that the ability to detect AI in this manner may have already be impossible and that at the time of writing enforcement will only catch the lazy and the desperate, and may be biased against non-native English writers (Liang et al 2023). This reinforces the previous conclusion that some forms of in-person assessment are necessitated. To be deliberately provocative, we acknowledge that we have used GPT-4 here to help with the writing process of this very document (explicitly the introduction). In order to meet the authorship criteria of the IOP and adhere to the requirements of the International Council of Medical Journal Editors (ICMJE) we also declare here that such use was to generate ideas and we have critically revised the work for intellectual content.

We make two further controversial points here. Firstly, it is now likely that much preceding literature on assessment must at minimum be regarded with more circumspect as it could potentially be the case that pre-November 2022 recommendations can make assessment more vulnerable to unauthorised use of LLMs in general and this raises deep questions about how we avoid such a scenario. We argue that only the above forms of assessment such as invigilated in-person examinations avoid this. We acknowledge the issues with these forms of assessment, but at this point it appears unavoidable.

Secondly, and very controversially, it may be the case that some “rote learning” could now be beneficial to test. For decades, the movement in the discipline (and every discipline) has been to veer away from memorisation of factual, and move to competency-based education (or at minimum outcome-based learning; Zaharia 2024), and from formal examination to authentic assessment (Palomba & Banta, 1999). However, an over-reliance on AI has the potential to de-skill future generations if students are not supported and assessed in their mastery of the underpinning discipline.

**Path 2: Integration**: A counterpoint to the argument that students must be assessed without the use of LLMs to ensure assessment validity is that these tools have the potential to be powerful learning tools, if embedded into teaching in an authentic manner (Walter 2024; see also Küchemann et al. 2024). As these tools become increasingly ubiquitous, students should be supported to engage with them appropriately. An extensive body of scholarly work (e.g. Beckingham et al., 2024) supports this 'embrace and adapt' strategy (JISC 2023). Core to a strategy where AI tools are embedded is the need to support students with critical literacies around their use (Ewen 2022) ensuring that students do not become uncritically reliant on them. This too presents a need to rethink what authentic assessment *is*, and how it might be adapted to reflect emerging workplace trends. As such, the use of AI tools becomes a crucial graduate skill (Beckingham et al., 2024), but one which must be embedded in a sound understanding of the discipline, as prompt engineering skills alone are insufficient (cf., Smith 2024; Wulff 2024).

In disciplines such as Physics, where problems are often (but not always) numerical and/or coding-based, where AI tools perform exceptionally well, the solution to assessment is very likely to be a balance between ensuring students have the necessary sound command of the field (perhaps through in-person assessment), which would allow them to expedite their work using AI tools where appropriate, identifying biases and weaknesses rather than uncritically accepting the output as fact. Our "maximally intelligent cheat" then becomes a "maximally intelligent user". In other disciplines, where assessments tend towards the more narrative and where rote learning may be less likely to be considered key to demonstrating *understanding* of the discipline, AI tools present both a challenge and an opportunity (Beckingham et al., 2024).

Finally, we asked GPT-4 itself what it considers to be problems it would struggle with. It cites highly complex non-linear systems (i.e., chaotic problems) that would exceed its capabilities, cutting-edge research problems, deep mathematic problems requiring advanced computation, anything in the lab that would require human "hands on", and problems requiring deep intuition on inter-disciplinary boundaries that could be ill-defined. When prompted to address if there's any problems humans are explicitly better at than AI, it reinforces the concept that problems requiring deep intuition and insight remain very squarely in the human domain, problems where humans navigate plentiful uncertainties in unexplored territories, physical experiments, interdisciplinary problems, and deep application of ethics and philosophy. There are arguably unsurprising, but we suspect these regimes will also narrow in the future but for now give a suggestion on what areas of coursework might still remain somewhat immune to unauthorised use of GPT-4: especially interdisciplinary working.

## 7. Conclusions.

We have presented work that shows GPT-4 is capable of tackling a physics degree. We note explicitly that it fails due to very specific circumstance: the inability to tackle in-person laboratory assessment, and in-person viva assessment.

We conclude that there is now a necessity to re-think and revise assessment practice in physics – and other disciplines – due to the challenge and opportunity provided by AI such as GPT-4. The use of invigilated in-person examinations, vivas, laboratory skills testing (or "performances" in other disciplines) can be used to ensure that students have the necessary mastery of the discipline to engage with AI tools intelligently, ethically and critically during other parts of their degree, and in their later employment. As educators, our task is to understand the tools available, the balance and integration between mastery and authenticity, set appropriate assessment tasks, and articulate clearly to our students the purposes of the tasks that we set.

The ability to detect AI in coursework may (or may not) already by impossible. We therefore suggest controversially that some amount of testing of mastery of disciplinary principles may be required to prevent de-skilling of future generations. On the other hand, it is probably the case that educators disagree and argue that testing such mastery is not controversial at all. Mixed opinions such as this serves as an indicator of how complex the future transformation in physics education (and other disciplines) is likely to be.

There are still cognitive tasks within the domain of physics where human beings will systematically outperform AI. These will narrow in the years ahead. For other disciplines, a similar analysis would be beneficial, to examine if the conclusions drawn here are more widely applicable and to determine if assessments are vulnerable in a systematic way.

Finally, it is instructive to note that Villalobos et al. (2022) suggest that high quality language data will run out for training purposes by 2026. Low quality might take until 2030 to fully harvest, and imaging data 2060. When all human output (to date) has been harvested, do we consider ourselves superior any longer? This says nothing of the possibility of artificial general intelligence either. We need to reform how we undertake assessment with celerity.

**Acknowledgments.**
KAP thanks colleagues in the Department of Physics at the University of Hull for cooperation over many years. In particular, Gareth Few has provided helpful conversations on this work. We also thank ViCE/PHEC where an earlier version of this work was presented by KAP in 2023 at the University of Durham for stimulating feedback and discussion.

KAP and LJM sincerely thank the two referees for providing feedback and encouragement that has improved this work, and the European Journal of Physics editors for their diligence.

**References.**